\documentclass[runningheads,fleqn]{cl2emult}
\usepackage{amsfonts}

\usepackage{graphicx}
\usepackage{graphics}
\usepackage{epsfig}
\usepackage{subeqnar}
\usepackage{multicol}
\usepackage{cropmark}
\usepackage{engi}
\usepackage{amsmath}

\begin{document}

\title{On the use of continuous wavelet analysis for modal identification}
\author{Pierre Argoul, Silvano Erlicher}

\toctitle{On the use of continuous wavelet analysis for modal
identification} % allows explicit linebreak for the table of content
\titlerunning{On the use of continuous wavelet analysis for modal
identification}
% allows abbreviation of title, if the full title is too long
% to fit in the running head
%

\authorrunning{Pierre Argoul and Silvano Erlicher}

\institute{Institut Navier - LAMI\\
         ENPC, 6-8 avenue Blaise Pascal,\\
         Cit\'{e} Descartes, Champs-sur-Marne,\\
         F-77455 Marne la Vall\'{e}e Cedex 2, France\\
         E-mail: argoul@lami.enpc.fr}

\maketitle

\begin{abstract}
This paper reviews two different uses of the continuous wavelet
transform for modal identification purposes. The properties of the
wavelet transform, mainly energetic, allow to emphasize or filter
the main information within measured signals and thus facilitate the
modal parameter identification especially when mechanical systems
exhibit modal coupling and/or relatively strong damping.
\end{abstract}

\section{Introduction}

The concept of wavelets in its present theoretical form was first
introduced in the 1970s by Jean Morlet and the team of the Marseille
Theoretical Physics Center in France working under the supervision
of Alex Grossmann. Grossmann and Morlet\cite{Grossman85} developed
then the geometrical formalism of the continuous wavelet transform.
Wavelets analysis has since become a popular tool for engineers. It
is today used by electrical engineers engaged for processing and
analyzing non-stationary signals and it has also found applications
in audio and video compression. Less than ten years ago, some
researchers proposed the use of wavelet analysis for modal
identification purposes. However, use of wavelet transforms in real
world mechanical engineering applications remains limited certainly
due to widespread ignorance of their properties.

\noindent This paper presents two ways of using the continuous
wavelet transform (CWT) for modal parameter identification according
to the type of measured responses : free decay or frequency response
functions (FRF).

\noindent In 1997, Staszewski \cite{Staszewski97} and Ruzzene
\cite{Ruzzene97} started to use wavelet analysis with the Morlet
wavelet for the processing of free decay responses, but the
identification of the mode shapes was not performed. Wavelet
transform allows to reach the time variation of instantaneous
amplitude and phase of each component within the signal and makes
the identification procedure of modal parameters much easier.
Compared to other mother wavelets, Morlet wavelet provides better
energy localizing and higher frequency resolution but has a
disadvantage: frequency-coordinate window shifts along frequency
axis with scaling while other wavelet types only expands the window.
To remedy it, Lardies et al. \cite{Lardies04} and Slavic et al.
\cite {Slavic03} preferred the Gabor wavelet. Argoul et al. \cite
{arg03a,arg03b} chose the Cauchy wavelet and Le et al. \cite{arg04}
established a complete modal identification procedure with
improvements for numerical implementation, especially for a correct
choice of the time-frequency localization. Edge effects are seen by
all the authors and some attempts to reduce their negative influence
on modal parameter identification are presented in \cite {Slavic03},
\cite{arg04} and \cite{Slavic04}.

\noindent More recently, Staszewski \cite{Staszewski98}, Bellizzi et
al. \cite {Bellizzi01} and Argoul et al. \cite{arg03a} adapted their
identification procedure in order to process free responses of
nonlinear systems. In \cite{arg03a}, the authors proposed four
instantaneous indicators, the discrepancy of which from linear case
facilitates the detection and characterization of the non-linear
behaviour of structures. In \cite{Bellizzi01}, an identification
procedure of the coupled non-linear modes is proposed and tested on
different types of non-linear elastic dynamic systems.

\noindent Typically, CWT is applied to time or spatial signals.
Processing frequency signals is unusual, but a first application was
proposed in \cite{arg97} and \cite{yin99}. Argoul showed in
\cite{arg97} that the weighted integral transform (WIT) previously
introduced by J\'{e}z\'{e}quel et al. \cite {arg86} for modal
identification purposes can be expressed by means of a wavelet
transform using a complex-valued mother wavelet close to the Cauchy
wavelet. The WIT can be applied to either the ratio of the time
derivative of the FRF\ over the FRF or the FRF itself. The
representation with CWT provided a better understanding of the
amplification effects of the WIT. When the FRF signal is strongly
perturbed by noise, its derivative is hard to obtain; thus, Yin et
al. \cite{yin99} then proposed a slightly modified integral
transform directly applied to the FRF and called the Singularities
Analysis Function (SAF) of order $n$. It allows the influence of the
FRF's poles to be emphasized and direct estimation of the eigen
frequencies, eigen modes, and modal damping ratios is performed from
the study of the extrema of the WIT.

\section{Theoretical background for the continuous wavelet analysis}

A wavelet expansion uses translations and dilations of an analyzing function
called the mother wavelet $\psi \in L^{1}(\mathbb{R})\cap L^{2}(\mathbb{R})$%
. For a continuous wavelet transform (CWT), the translation and
dilation parameters: $b$ and $a$ respectively, vary continuously. In
other words,
the CWT uses shifted and scaled copies of $\psi (x)$ : $\psi _{b,a}(x)=\frac{%
1}{a}\psi (\frac{x-b}{a})$ whose $L^{1}(\mathbb{R})$ norms ($\left\|
.\right\| _{1}$) are independent of $a$. In the following, $\psi
(x)$ is assumed to be a smooth function, whose modulus of Fourier
transform is peaked at a particular frequency $\Omega _{0}$ called
the ''central'' frequency. The variable $x$ may represent either
time or frequency; when necessary, the time and circular frequency
variables will be referred to $t$ and to $\omega $ respectively. The
CWT of a function $u(x)\in L^{2}(\mathbb{R})$
can then be defined by the inner product between $u(x)$ and $\psi _{b,a}(x)$%
\begin{equation}
T_{\psi }[u](b,a)=\left\langle u,\psi _{b,a}\right\rangle =\frac{1}{a}%
\int_{-\infty }^{+\infty }u(x)\,\overline{\psi }\left( \frac{x-b}{a}\right)
dx  \label{deficwt}
\end{equation}
where $\overline{\psi }(.)$\ is the complex conjugate of $\psi (.)$. Using
Parseval's identity, Eq. (\ref{deficwt}) becomes
\begin{equation}
T_{\psi }[u](b,a)=\frac{1}{2\pi }\left\langle \widehat{u},\widehat{\psi }%
_{b,a}\right\rangle =\frac{1}{2\pi }\int_{-\infty }^{+\infty }\widehat{u}%
(\Omega )\,\overline{\widehat{\psi }}(a\Omega )\,e^{i\Omega b}\,d\Omega
\label{cwtdual}
\end{equation}
where $\widehat{u}(\Omega )$, $\widehat{\psi }_{b,a}(\Omega )$ and $\widehat{%
\psi }(\Omega )$ are respectively the Fourier transform (FT) of $u(x)$, $%
\psi _{b,a}(x)$ and $\psi (x)$; for instance for $u(x)$ : $\widehat{u}%
(\Omega )=\int_{-\infty }^{+\infty }u(x)\,e^{-i\Omega x}dx$.

\noindent Moreover, when $\psi $ and $u$ are continuous and piece-wise
differentiable, and $\dot{\psi}$ is square and absolutely integrable, and $%
\dot{u}$ is of finite energy, the CWT of $\dot{u}$ with $\psi $ is linked to
the CWT of $u$ with $\dot{\psi}$ :
\begin{equation}
T_{\psi }[\dot{u}](b,a)=-\frac{1}{a}T_{\dot{\psi}}[u](b,a)  \notag
\end{equation}
\noindent From \cite{chui91}, it can be seen that the CWT at point $%
(b,\Omega =\frac{\Omega _{\psi }}{a})$ picks up information about
$u(x)$, mostly from the localization domain $D(b,\Omega
=\frac{\Omega _{\psi }}{a})$ of the CWT, defined as {\small
\begin{equation} D(b,\Omega =\frac{\Omega _{\psi }}{a})=\lbrack
b+ax_{\psi }-a\Delta x_{\psi },b+ax_{\psi }+a\Delta x_{\psi
}\,]\times \lbrack \frac{\Omega _{\psi }}{a}-\frac{\Delta \Omega _{\psi }}{a}%
\,,\frac{\Omega _{\psi }}{a}+\frac{\Delta \Omega _{\psi }}{a}]
\label{timefrequencydomain}
\end{equation}}
\noindent where $x_{\psi }$ and $\Delta x_{\psi }$ are called centre
and radius
of $\psi ,$ stated in terms of root mean squares : $x_{\psi }=\frac{1}{%
\left\| \psi \right\| _{2}^{2}}\int_{-\infty }^{+\infty }x\,\,\left| \psi
(x)\right| ^{2}\,dx$ and \newline
\noindent $\Delta x_{\psi }=\frac{1}{\left\| \psi \right\| _{2}}\sqrt{%
\int_{-\infty }^{+\infty }\left( x-x_{\psi }\right) ^{2}\,\left| \psi
(x)\right| ^{2}\,dx}$, and similar definitions hold on for the frequency centre $%
\Omega _{\psi }$ and the radius $\Delta \Omega _{\psi }$ of $\widehat{%
\psi }$. The area of $D$ is constant and equal to four times the uncertainty : $%
4\Delta x_{\psi }\Delta \Omega _{\psi }=4\mu (\psi )$. The
Heisenberg uncertainty principle states that this area has to be
greater than $2$.

\noindent Referring to the conventional frequency analysis of
constant-$Q$\ filters, the parameter $Q$ defined as the ratio of the
frequency centre $\Omega _{\psi }$ to the frequency bandwidth
($2\Delta \Omega _{\psi }$):
\begin{equation}
Q\,=\,\frac{\frac{\Omega _{\psi }}{a}}{2\frac{\Delta \Omega _{\psi }}{a}}%
\,=\,\frac{\Omega _{\psi }}{2\Delta \Omega _{\psi }},  \label{equ:Q}
\end{equation}
is introduced in \cite{arg04} to compare different mother wavelets
and to characterize the quality of the CWT. $Q$\ is independent of
$a$.

\noindent The notion of spectral density can be easily extended to the CWT
(see \cite{carmona98} for others properties such as linearity, admissibility
and signal reconstruction, etc.). From the Parseval theorem applied to Eq.
\ref{cwtdual}, it follows that
\begin{equation}
\int_{-\infty }^{+\infty }\left| T_{\psi }[u](b,a)\right| ^{2}\,db=\frac{1}{%
2\pi }\int_{-\infty }^{+\infty }\left| \widehat{u}(k)\right| ^{2}\,\left|
\overline{\widehat{\psi }}(ak)\right| ^{2}\,dk  \notag
\end{equation}
and it leads to the ''energy conservation'' property of the CWT expressing
that if $u\in L^{2}(\mathbb{R})$, then $T_{\psi }\in L^{2}(\mathbb{R\times R}%
_{+}^{\ast },db\frac{da}{a})$, and $\left\| T_{\psi }[u]\right\|
^{2}=C_{\psi }\,\left\| u\right\| ^{2}=\frac{C_{\psi }}{2\,\pi }\left\|
\widehat{u}\right\| ^{2}$. Finally, after changing the integration variable: $a=%
\frac{\Omega _{0}}{\Omega }$ and because $\widehat{u}(\Omega )$ is hermitian
($u(x)$ being real valued), one gets
\begin{equation}
\int_{0}^{+\infty }\int_{-\infty }^{+\infty }\left| T_{\psi }[u](b,\frac{%
\Omega _{o}}{\Omega })\right| ^{2}db\frac{d\Omega }{\Omega }=\frac{C_{\psi }%
}{\,\pi }\int_{0}^{+\infty }\left| \widehat{u}(\Omega )\right| ^{2}d\Omega
\label{energytoc1}
\end{equation}
where $C_{\psi }=\int_{-\infty }^{+\infty }$ $\frac{\left| \widehat{\psi }%
(\omega )\right| ^{2}}{\omega }d\omega <\infty $.

\noindent This allows to define a local wavelet spectrum: $E_{u,CWT}(\Omega
,b)=\frac{1}{2C_{\psi }\,\Omega }\left| T_{\psi }[u](b,\frac{\Omega _{o}}{%
\Omega })\right| ^{2}$ and a mean wavelet spectrum: $E_{u,CWT}(\Omega )=$\ $%
\int_{-\infty }^{+\infty }E_{u,CWT}(\Omega ,b)db$; and Eq.
\ref{energytoc1} can be rewritten: $\int_{0}^{+\infty }\left[ E_{u,CWT}(\Omega )-\frac{1}{2\pi }\left| \widehat{u%
}(\Omega )\right| ^{2}\right] \,d\Omega =0$.

\noindent When the mother wavelet $\psi $ is progressive (i.e. it
belongs to the complex Hardy space $\widehat{\psi }(\Omega
)=0:\mbox{for}:\Omega \leq 0$), the CWT of a real-valued signal
$u$ is related to the CWT of its analytical signal $Z_{u}$ (see.
\cite{carmona98})
\begin{equation}
T_{\psi }[u](b,a)=\frac{1}{2}T_{\psi }[Z_{u}](b,a)  \label{cwtanalytical}
\end{equation}
From \ref{timefrequencydomain}, the CWT has sharp frequency
localization at low frequencies, and sharp time localization at high
frequencies. Thanks to a set of mother wavelets depending on
one(two) parameter(s), the desired time or frequency localization
can be yet obtained by modifying its(their) value(s). In
\cite{arg04}, two complex valued mother wavelets were analyzed :
Gabor and Cauchy.\ Morlet wavelet is a complex sine wave localized
with a Gaussian envelope and the Gabor wavelet function is a
modified Morlet wavelet with a parameter controlling its shape.
Cauchy wavelet $\psi _{n}$ of $n$ order for $n\geq 1$, is defined
by: {\small
\begin{equation}
 \psi _{n}(x)=\left( \frac{i}{x+i}\right)
^{n+1}=\left( \frac{1}{1-ix}\right) ^{n+1} =%
\left[ \frac{1}{\sqrt{x^{2}+1}}\right] ^{n+1}e^{i\,(n+1)%
Arctg(x)}.  \label{ondemere}
\end{equation} }

\noindent The main characteristics of $\psi _{n}(x)$ follow: its FT:
$\widehat{\psi }_{n}(\Omega )=\frac{2\pi \Omega ^{n}e^{-\Omega
}}{n!}\,\Theta (\Omega )$ where $\Theta (\Omega )$ is the Heaviside
function ($\psi _{n}$ is progressive), its central frequency:
$\Omega _{0_{n}}=n$, its L2 norm: $\left\| \psi _{n}\right\|
_{2}^{2}=\frac{(2n)!}{2^{2n}(n!)^{2}}\pi $, its x- and frequency
centres: $x_{\psi }=0$ and $\Omega _{\psi }=n+\frac{1}{2}$, its
radius: $\Delta x_{\psi }=\sqrt{\frac{1}{2n-1}}$, its frequency
radius:\ $\Delta \Omega _{\psi }=\frac{\sqrt{2n+1}}{2}$, its
uncertainty: $\mu _{\psi }=\frac{1}{2}\sqrt{1+\frac{2}{2n-1}}$, its
quality factor: $Q=\frac{n+\frac{1}{2}}{\sqrt{2n+1}}$ and its
admissibility factor:
$C_{\psi }=4\pi ^{2}\frac{1}{2^{2n}}\frac{(2n-1)!}{(n!)^{2}}=\frac{2\pi }{n%
}\left\| \psi _{n}\right\| _{2}^{2}$. The use of the Cauchy wavelet
is legitimate when $Q$ is less than $5/\sqrt{2}$; as $Q$ increases,
both wavelets give close results, a little better with Morlet
wavelet due to its excellent time-frequency localization, and behave
similarly in the both time and frequency domains when $Q$ tends
toward infinity.

\noindent Some numerical and practical aspects for the computation
of the CWT are detailed in \cite{arg04}, especially when applied to
signals with time bounded support.\ Let us note $D_{u}=\left[
0,L\right] \times \left[ 0,2\pi f_{Nyq}\right] $ the validity domain
of the discretized signal in the
$(b,\Omega )$ plane where $%
f_{Nyq} $ is the Nyquist frequency ($f_{Nyq}=\frac{1}{2T}$ with
sampling period $T$). The edge-effect problem due to the finite
length ($L$) and to the discretization of measured data record and
to the nature of the CWT (convolution product) is tackled. An
extended domain $D_{ext}(b,\Omega)\supset D(b,\Omega)$ where $\Omega
=\frac{\Omega _{\psi }}{a})$, is then proposed to take into account
the decreasing properties of $\psi $ and $\widehat{\psi }$\ by
introducing two real positive coefficients $c_{x}$ and $c_{\Omega }$: $%
{\footnotesize D_{ext}(b,\Omega =\frac{\Omega _{\psi
}}{a})=[b+ax_{\psi }-a\,c_{x}\Delta x_{\psi },b+ax_{\psi
}+ac_{x}\Delta x_{\psi }\,]\times \lbrack \frac{\Omega _{\psi
}}{a}-c_{\Omega }\frac{\Delta \Omega _{\psi }}{a}}$ ,
${\footnotesize
\frac{\Omega _{\psi }}{a}+c_{\Omega }\frac{\Delta \Omega
_{\psi }}{a}]}$.  Forcing $%
D_{ext} $ to be included into $D_{u}$ leads the authors to define a
region in the $(b,\Omega )$ plane where the edge effect can be
neglected; this
region is delimited by two hyperbolae whose equations are: $\Omega =\frac{2}{b%
}c_{x}Q\mu _{\psi }$ and $\Omega =\frac{2}{L-b}c_{x}Q\mu _{\psi }$
and two horizontal lines whose equations are: $\Omega =0$ and
$\Omega =\frac{2\pi f_{Nyq}}{1+c_{\Omega }\frac{1}{2Q}}$. Let us
consider a frequency $\Omega _{j}$ for which $\widehat{u}(\Omega )$\
exhibits a peak and introduce a frequency discrepancy $d\Omega _{j}$
(for example, the distance between two successive peaks of
$\widehat{u}(\Omega)$). From the intersection of the two hyperbolae
at the point $(b={\footnotesize \frac{L}{2}},\Omega={\footnotesize
\Omega_{j}})$ and by imposing the
frequency localization along the straight line: $a={\footnotesize \frac{\Omega _{0}}{%
\Omega _{j}}}$, to be included into $\left[ \Omega _{j}-d\Omega
_{j},\Omega _{j}+d\Omega _{j}\right] $, some upper and lower bounds
are found for $Q$: $c_{\Omega }\frac{\Omega _{j}}{2d\Omega _{j}}$\
$\leq Q\leq \frac{L\,\Omega _{j}}{2c_{x}}$ and finally, the authors
proposed $c_{x}=c_{\Omega }\simeq5$.

\section{Modal analysis and modal identification with CWT}

Experimental identification of structural dynamics models is usually
based on the modal analysis approach. One basic assumption
underlying modal analysis is that the behaviour of the structure is
linear and time invariant during the test. Modal analysis and
identification involve the theory of linear time-invariant
conservative and non-conservative dynamical systems. In this theory,
the normal modes are of fundamental importance because they allow to
uncouple the governing equations of motion. Also, they can be used
to evaluate the free or forced dynamic responses for arbitrary sets
of initial conditions. Modal analysis of a structure is performed by
making use of the principle of linear superposition that expresses
the system response as a sum of modal responses.

\noindent For linear MDOF\ systems with $N$ degrees of freedom, the transfer
function $\mathbb{H}_{ij}(p)$ of receptance type is defined as the Laplace
transform of the displacement at point $j$ \ to the impulse unit applied at
point $i$ and it can be expressed as a sum of simple rational fractions
\begin{equation}
\mathbb{H}_{ij}(p)=\sum_{r=1}^{N}\left( \frac{\left( A_{r}\right) _{ij}}{%
p-p_{r}}+\frac{\left( \overline{A_{r}}\right) _{ij}}{p-\overline{p_{r}}}%
\right)  \label{fonctiontransfert}
\end{equation}
where $\left( \overline{A_{r}}\right) _{ij}$ and $\overline{p_{r}}$
are respectively the conjugate of the residues $\left( A_{r}\right)
_{ij}$ and of
the poles $p_{r}$ of $\mathbb{H}_{ij}(p)$. When the system is stable, $%
p_{r}=-\xi _{r}\,\omega _{r}+i\widetilde{\omega }_{r}$ where $\omega _{r}$, $%
\widetilde{\omega }_{r}$ are the undamped and the damped vibration
angular frequencies and $\xi _{r}$ is the damping ratio for the mode
$r$. $\left( A_{r}\right) _{ij}$ can also be expressed from the
complex modes $\chi _{r}$ by $\left( A_{r}\right) _{ij}=\frac{\left(
\chi _{r}\right) _{i}\,\left(
\chi _{r}\right) _{j}}{\gamma _{r}}$ where $\gamma _{r}=\chi _{r}^{T}%
\underline{\underline{C}}\chi _{r}+2p_{r}\chi _{r}^{T}\underline{\underline{M%
}}\chi _{r}$, $\underline{\underline{M}}$ and
$\underline{\underline{C}}$ being respectively the mass and viscous
damping matrices. Moreover, the FRF of receptance type
$H_{ij}(\omega )$ can be usually related to $\mathbb{H}_{ij}(p)$ by:
$H_{ij}(\omega )=\mathbb{H}_{ij}(p=i\omega)$, leading to
\begin{equation}
H_{ij}(\omega )=\sum_{r=1}^{N}\left( \frac{%
-i\left( A_{r}\right) _{ij}}{\left( \omega -\widetilde{\omega }_{r}\right)
-i\xi _{r}\,\omega _{r}}+\frac{-i\left( \overline{A_{r}}\right) _{ij}}{%
\left( \omega +\widetilde{\omega }_{r}\right) -i\xi _{r}\,\omega _{r}}\right)
\label{FRF2}
\end{equation}

\noindent The free responses at point $j$ in terms of displacements $%
u_{j}^{_{(free)}}(t)$ can be expressed according to the modal basis
of complex modes: $\chi _{jr}=\eta _{jr}+i\,\kappa _{jr}$ ($r$ being
the mode number, $1\leq r\leq N$)
\begin{equation}
u_{j}^{_{(free)}}(t)=\sum_{r=1}^{N} A_{rj}(t) \cos \left( \Phi
_{rj}(t) \right) \label{free1}
\end{equation}
where $A_{rj}(t)=\left| \chi _{jr}\right| \rho
_{r}\,e^{-\xi _{r}\omega _{r}t}$ and $\Phi _{rj}(t)=\widetilde{\omega }%
_{r}t-\varphi _{r}+\arctan \left( \frac{\kappa _{jr}}{\eta
_{jr}}\right) $. $\rho _{r}$ and $\varphi _{r}$ are defined from
initial displacement and velocity of mode $r$ (cf. \cite{gerad96}).

\noindent In the case of proportional viscous damping (Basile
assumption), introducing the real eigenvectors $\Psi _{r}$ of mode $%
r$ for the associated conservative system, and replacing the
residues $\left( A_{r}\right) _{ij}$ with $\frac{\left(
\Psi _{r}\right) _{i}\,\left( \Psi _{r}\right) _{j}}{2i\,\widetilde{\omega }%
_{r}m_{r}}$ in Eq. \ref{FRF2} lead to
\begin{equation}
H_{ij}(\omega )=\sum_{r=1}^{N}\frac{\left( \Psi _{r}\right) _{i}\,\left(
\Psi _{r}\right) _{j}}{m_{r}\left( \omega _{r}^{2}-\omega ^{2}+2i\,\xi
_{r}\,\omega _{r}\omega \right) }  \label{FRF}
\end{equation}
 \noindent Moreover, in Eq. \ref{free1}, $\kappa _{jr}=0$;
thus $\left| \chi _{jr}\right| $ can be replaced by $\left| \left(
\Psi _{r}\right)
_{j}\right| $ and the term $\arctan \left( \frac{\kappa _{jr}}{\eta _{jr}}%
\right) $ can be equal to $0$ or $\pi $ according to the sign of the
real part $\eta _{jr}$.

\subsection{Modal identification using free decay responses}

The processed signals $u_{j}^{_{(free)}}(t)$, whose expression is
given in Eq. \ref{free1}, can be considered under the assumption of
weak damping
($\xi _{r}\ll \frac{1}{\sqrt{2}}$), as a sum of $N$ modal components: $%
A_{rj}(t)\,\cos (\Phi _{rj}(t))$\ consisting in asymptotic signals.
The approximation of asymptotic signal means that the oscillations
resulting from the phase term: $\Phi _{rj}(t)$ are much faster than
the variation coming from the amplitude term: $A_{rj}(t)$; it
entails that the analytical signal associated with
$A_{rj}(t)\,e^{i\Phi _{rj}(t)}$ can be approximated by:
$A_{rj}(t)\,e^{i\Phi _{rj}(t)}$. Therefore, Eq.
(\ref{cwtanalytical})
and the linearity of the CWT entail that: $%
T_{\psi }[u_{j}^{_{(free)}}](b,a)=\frac{1}{2}\sum_{r=1}^{N}T_{\psi
}[A_{rj}(t)\,e^{i\Phi _{rj}(t)}](b,a)$. Taking the Taylor expansion
of each amplitude term of this sum, it follows that
\begin{equation}
T_{\psi }[u_{j}^{_{(free)}}](b,a)=\frac{1}{2}\sum_{r=1}^{N}\left(
A_{rj}(b)\,e^{i\Phi _{rj}(b)}\,\overline{\widehat{\psi }(a\,\widetilde{%
\omega }_{r})}+R_{rj}(b,a)\right). \label{cwtasymptoticsignal}
\end{equation}

\begin{figure}[ht]
\begin{center}
\includegraphics*[width=75mm]{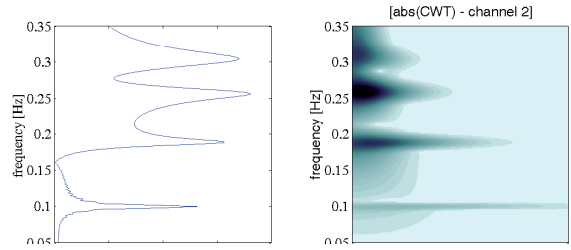}
\end{center}
\caption{Modulus of FT and of the CWT for the 4DoF system ($Q=18$) }
\label{moduli4dof}
\end{figure}
\begin{figure}[ht]
\begin{center}
\includegraphics*[width=75mm]{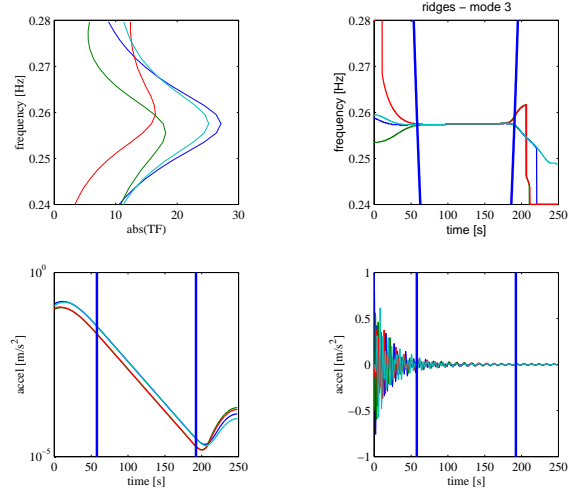}
\end{center}
\caption{Modulus of FT and of CWT around the 3rd mode for all
channels. Logarithm of the modulus of the CWT ($L=250$s, $T=0.175$s,
$c_{t}=c_{\Omega}=4.7$, $Q=20$).} \label{mode3}
\end{figure}

\noindent Assuming that each $R_{rj}(b,a)$\ is small enough to be
neglected, $\left| T_{\psi }[u_{j}^{_{(free)}}](b,a)\right| $\ is
peaked in the $(b,a)$ plane near straight lines of equation:
$a=\frac{\Omega _{0}}{\Phi _{rj}^{\prime }(b)}=\frac{\Omega
_{0}}{\widetilde{\omega }_{r}}$. When each modal pseudo-pulsation
$\widetilde{\omega }_{r}$ is far enough
from the others, this straight line's equation easily provides an approximation of $\widetilde{%
\omega }_{r}$. Eq. \ref{cwtasymptoticsignal} implies that the CWT of
asymptotic signals has a tendency to concentrate near a series of
curves in the time-frequency plane called ridges, which are directly
linked to the amplitude and phase of each component of the measured
signal. In \cite{arg04}, a complete modal identification procedure
for natural frequencies, viscous damping ratios and mode shapes, is
given in the case of viscous proportional damping and applied to a
numerical case consisting in the free decay responses of a
mass-spring-damper system with four degrees of freedom (4-DoF). The
procedure is applied here to the free responses of a 4-DoF system
whose natural frequencies are: $f_{1}=0.0984$ Hz, $f_{2}=0.1871$ Hz,
$f_{3}=0.2575$ Hz and $f_{4}=0.3027$ Hz and modal damping ratios:
$\xi_{1}=0.0124$, $\xi _{2}=0.0235$, $\xi _{3}=0.0324$, $\xi
_{4}=0.0380$. Fig. \ref{moduli4dof} shows the FT and the CWT of the
displacement of the second mass for which the four eigenfrequencies
are clearly visible. Fig. \ref{mode3} presents for the third mode
and for the four masses, the modulus of the CWT and its logarithm
that will allow to estimate the value of the damping ratio $\xi
_{3}$ (cf. \cite{arg04}). The edge effect is delimited by two
hyperbolae in the time frequency plane; $Q$ being chosen to $20$
($15.3\leq Q\leq 43.1$). Identified values are very close to the
exact ones and identification errors are negligible inside the
domain $D_{ext}$. This procedure was also applied in \cite{arg03b}
to a set of accelerometric responses of modern buildings submitted
to non destructive shocks. From the CWT of measured responses, the
ridge and the corresponding amplitude and phase terms for the two
first modes were extracted. For the first instantaneous frequency
$\widetilde{\omega }_{1}$, the processing revealed a slight increase
of $\widetilde{\omega }_{1}$ just after the shock. The origin of
this non linear effect was attributed to the non-linear behaviour of
the soil-structure interaction. Other preliminary results can be
found in \cite{arg03a} where instantaneous indicators based on CWT
are computed from the accelerometric responses of a non-linear beam
to an impact force. A Duffing non-linearity effect was then
identified thanks to the first and the super-harmonic components.

\subsection{Modal identification using FRF functions}

The first definition of the SAF of order $n$ ($n\in\mathbb{N}$) uses
the derivative of a transfer function $\mathbb{H}(p)$ of the linear
mechanical system, where $p=a+ib$ :
\begin{equation}
\mathbf{H}_{n}(b,a)=(-1)^{n}2\,\pi \,a^{\frac{n+1}{2}}\frac{d^{n}}{dp^{n}}%
\left\{ \mathbb{H}(p)\right\}   \label{FAS1}
\end{equation}
$\mathbf{H}_{n}(b,a)$ can also be expressed as the FT of the impulse
response function $h(t)$\ filtered by: $2\pi a^{\frac{n+1}{2}%
}t^{n}e^{-at}\Theta (t)$ where $\Theta (t)$ is the Heaviside function :
\begin{equation}
\mathbf{H}_{n}(b,a)=2\pi \,a^{\frac{n+1}{2}}\int_{\mathbb{R}%
^{+}}t^{n}e^{-at}\,h(t)\,e^{-ibt}\,dt  \label{FAS2}
\end{equation}
Finally, $\mathbf{H}_{n}(b,a)$ is equal to the CWT of the FRF
$H(\omega )$ with a mother wavelet being the conjugate
of the Cauchy wavelet of order $n$ multiplied by $\frac{n!}{a^{%
\frac{n-1}{2}}}$ :
\begin{equation}
\mathbf{H}_{n}(b,a)=\frac{n!}{a^{\frac{n+1}{2}}}T_{\overline{\psi}_{n}}[H](b,a)=%
\frac{n!}{a^{\frac{n+1}{2}}}\int_{\mathbb{R}}H(\omega )\psi%
_{n}\left( \frac{\omega -b}{a}\right) \,d\omega   \label{FAS3}
\end{equation}

\noindent Let us now consider a transfer function restricted to only
one term $F_{\lambda ,\omega _{0}}(\omega )$\ of the sum given in
(\ref{FRF2}) and perturbed by a Gaussian white noise $n(\omega )$ so
that : $H(\omega )=F_{\lambda ,\omega _{0}}(\omega )+n(\omega
)=\frac{-B\,i}{\left( \omega -\omega _{0}\right) -i\lambda
}+n(\omega )$. Let us introduce the coefficient $\rho ^{2}$ defined
as the ratio of the square absolute value of the CWT of the pure
signal upon the common variance of the CWT\ of the noise itself:
$\rho ^{2}=\frac{\left| T_{\psi }[H](b,a)\right| ^{2}}{%
E\left\{ \left| T_{\psi }[n](b,a)\right| ^{2}\right\} }=\frac{\left|
\left\langle F_{\lambda ,\omega _{0}}\,,\,\psi _{b,a}\right\rangle
\right| ^{2}}{E\left\{ \left| \left\langle n\,,\,\psi
_{b,a}\right\rangle \right| ^{2}\right\} }$, it looks like a signal
to noise ratio.

\noindent The use of the Cauchy-Schwarz inequality gives that :
$\rho ^{2}\leq \left\| F_{\lambda
,\omega _{0}}\right\| ^{2}$ and the maximum of $\rho ^{2}$\ is reached for $%
(b,a)=(\omega _{0},\lambda )$ when $\psi _{b,a}(\omega
)=K\,F_{a,b}(\omega )=K\frac{-B\,i}{\left( \omega -b\right) -ia}$
where $K$ is a constant. In conclusion, $\psi (\omega )=K\,B\frac{-\,i}{\omega -i}=K\,B\,\overline{\psi}%
_{0}(\omega )$; the analyzing functions $\psi (\omega ) $ which
allow the minimization of the noise effect on the peaks of
$\left|T_{\psi }[H](b,a)\right|$ are proportional to
$\overline{\psi}_{0}(s)$ defined in Eq. \ref{ondemere}.

\noindent The above definitions of the SAF of order $n$
($n\in\mathbb{N}^{*}$) allow to better understand its effect when
applied to discrete causal linear mechanical systems. In Eq.
\ref{FAS1}, the successive derivatives of a rational fraction make
the degree of the denominator increase and cause an amplification of
the pole's effect. So, when Eq. \ref{fonctiontransfert} is
introduced in Eq. \ref{FAS1},\ the SAF\ of order $n$ becomes
\begin{equation}
\mathbf{H}_{n}(b,a)=2\pi \,n!\,a^{\frac{n+1}{2}}\sum_{r=1}^{N}\left( \frac{%
A_{r}}{\left( a+ib-p_{r}\right) ^{n+1}}+\frac{\overline{A_{r}}}{\left( a+ib-%
\overline{p_{r}}\right) ^{n+1}}\right)  \label{SAFfiltre}
\end{equation}
The absolute value of $\mathbf{H}_{n}(b,a)$ exhibits $2N$ maxima
located symmetrically from the vertical axis in the phase plane. The
effect of parameter $n$ is to facilitate the
 modal identification, especially when the
structure has neighboring poles. As soon as $n$ is large enough, the
coordinates ($b_{max},a_{max}$) of these $2N$ maxima allow the
estimation of the real and imaginary parts, respectively $\Re
\left\{ p_{r}\right\} $ and $\Im \left\{ p_{r}\right\} $, of each of
the $2N$ conjugate poles of the system :
\begin{equation}
\Re \left\{ p_{r}\right\}=-\xi _{r}\,\omega _{r} \approx
-a_{\max_{r}}\,\,\,\,\,\,\ \ \ \,\,\,\,\,\,\,\text{and}\,\,\,\,\,\,\
\ \ \,\,\,\,\,\,\,\Im \left\{ p_{r}\right\}=\widetilde{\omega }_{r}
\approx b_{\max_{r}}
\end{equation}
The residue $A_{r}$ can be then estimated from the
complex valued amplitude of the extremum called $\mathbf{H}_{n}^{\max_{r}}=%
\mathbf{H}_{n}^{\max }(a_{\max_{r}},b_{\max_{r}})$
\begin{equation}
A_{r}\approx \frac{2^{n}\left( a_{\max_{r}}\right) ^{\frac{n+1}{2}}}{\pi \,n!%
}\mathbf{H}_{n}^{\max_{r}}
\end{equation}
\noindent As $n$ increases, the estimation of the extrema is better
for signals without noise; but for noisy signals, noise effects are
also increased. Recently, Yin et al. \cite{yin04} proposed the use
of analyzing wavelets which are the conjugate of Cauchy wavelets in
which $n$ is replaced by a positive real number. Applications both
to numerical and experimental data showed the efficiency of this
technique using SAF \cite{arg99}. It was applied to the FRFs
obtained with H1 estimators, of a test structure designed and build
by ONERA\ for the GARTEUR-SM-AG19 Group. This structure was made of
two aluminum sub-structures simulating wings/drum and fuselage/tail.
Finally, the results obtained with SAF were very similar to those
obtained with the broadband MIMO modal identification techniques
provided by the MATLAB\ toolbox IDRC. \vspace{0.25cm}

 \noindent In
conclusion, the pre-processing of measured signals (free decay
responses or FRFs) with CWT proved to be an efficient tool for modal
identification.

%INDEX%%%%%%%%%%%%%%%%%%%%%%%%%%%%%%%%%%%%%%%%%%%%%%%%%%%%%%%%%%%%%%%
%\clearpage\addcontentsline{toc}{section}{Index} \flushbottom\printindex
%%%%%%

\end{document}